\documentstyle[prl,aps]{revtex}
\begin{document}

\title{Reducing Quantum Errors and Improving Large Scale Quantum Cryptography}

\author{Tal Mor$^{(1)}$}

\address{(1) Physics Department, Technion, Haifa
32000, Israel;}

\date{\today}

\maketitle

\begin{abstract}

Noise causes severe difficulties in implementing quantum 
computing and quantum cryptography.
Several schemes have been suggested to reduce this problem, 
mainly focusing on quantum computation.
Motivated by quantum cryptography, we suggest a coding which
uses $N$ quantum bits ($N=n^2$) to encode one quantum bit, and reduces 
the error exponentially with $n$. Our result suggests the possibility
of distributing a secure key over very long distances, and maintaining
quantum states for very long times. It also provides a new 
quantum privacy amplification against a strong adversary.

\end{abstract}


%\begin{multicols}{2}
\twocolumn

The ability to correct errors in a quantum bit (qubit) 
is crucial to the success of quantum computing,
and it is very important to the success of quantum cryptography. 
Motivated by quantum computing, 
Shor~\cite{Shor1} shows that quantum errors can be corrected
(in some analogy to classical error correction~\cite{MS}).
The many works which follow Shor's idea focus 
on improving his result~\cite{BDSW,LMPZ,Steane1,CS} to better fit 
the requirements of quantum computing, or to provide a 
better understanding of the properties of the error-correction 
codes (the previous works and also~\cite{Steane2,EM,Shor2}).  
In this work we apply this idea to 
quantum cryptography, where reducing the error rate is the main aim.
We emphasize the properties of quantum error-correction which are
relevant to quantum cryptography, and we
show that quantum cryptography
can be tremendously improved
using a simple generalization of Shor's scheme.

Quantum cryptography~\cite{BB84}
has already taken some 
promising experimental 
steps~\cite{Gisin}, and, to certain limits, it can work 
without involving the complications added by error-correction
(or more precisely, error-reduction) schemes.  
In reality, there is noise due to preparing, transmitting
and receiving the quantum states, and 
practical protocols deal with small error-rates. 
However, the noise still causes a severe problem due to the 
combination of the following two reasons:
(1) The information available to an eavesdropper (Eve) on a single bit
depends on the error-rate which the legitimate users 
(Alice and Bob) accept~\cite{BMS,BM},
and reducing this error-rate reduces Eve's information on the final key.
Moreover, security analysis (e.g.,~\cite{BM})
is restricted to small
error rates, and removing this restriction might make the analysis 
(technically) impossible.
(2i) Existing error-rates do not
allow key distribution over long distances, 
due to error accumulation over distance.
This is the main problem of practical
quantum key distribution, and 
currently~\cite{Gisin} the best existing systems
distribute a key to distances of up to 30~km.
The scheme we present here might enable
one to increase this distance significantly,
suggesting that a lot of effort should be spent on this direction.
(2ii) Some quantum cryptographic schemes~\cite{GV,BHM} 
use quantum memory instead
of (or in addition to) quantum channels. Reducing the errors
in such schemes is important since it allows keeping the states unchanged
for a desirable time.
(This was partly suggested in~\cite{VGW};
a discussion of related works is done in the concluding paragraph.)
Moreover, the quantum cryptographic network~\cite{BHM} which allows
communication between any two users (while using 
no quantum channels between them) already uses 
the same experimental ingredients
as error-reduction schemes. 
Therefore, improving quantum cryptography in the future,
using error-reduction schemes, 
might make such a scheme favorite.

The error-reduction 
scheme we suggest here allows, in principle, to reduce the noise in
a transmission channel or in a quantum memory
to any desirable level.
This result is important (from a theoretical point of view) for implementing
a ``quantum privacy amplification'' scheme, as an 
alternative to another quantum privacy amplification scheme~\cite{England}
which is based on purification of singlet-pairs~\cite{purification}.
Such schemes provide a promising direction for proving the ultimate
security of quantum cryptography,
as an alternative to the more practical approach of~\cite{BM}.

Classical error-correction is based 
on redundant encoding which uses more than one bit (on average) to  
encode one bit.
The simplest scheme is the $1\rightarrow 3$ repetition code in which
each bit is repeated three times in the encoding, and a majority
vote is chosen for decoding.
In this case, if a single bit contains an error with probability $p$
(where $p$ is small), and $p_l={3 \choose l} p^l (1-p)^{3-l}$ 
is the probability of having 
exactly $l$ errors, then 
the probability to have a remainder error (the probability of having two
or three errors in three bits) is 
$ P= p_2 + p_3 = 3 (1-p) p^2 + p^3 = 3 p^2 - 2 p^3 $. 
One must keep in mind that this result is true {\em on average}, but
in case we know that one error was identified and corrected
[which happens with probability $p_2 + p_1 = 3 p^2 (1-p)+3 p (1-p)^2$]
the probability of having a remainder error is exactly $p_2/(p_1+p_2)=p$,  
and we gain 
no error reduction at all!

The analogous quantum error-correction~\cite{Shor1}
uses 9 qubits to encode a single qubit 
(to perfectly correct a single error) using the following procedure:
a $1\rightarrow 3$ repetition code 
in the $z$ basis
$  |0\rangle \rightarrow |000\rangle $ and 
$  |1\rangle \rightarrow |111\rangle $
(where $|000\rangle$ stands for the tensor product
$|0\rangle |0\rangle |0\rangle$
of three qubits);
a transformation to the $x$ basis
$|0\rangle \rightarrow(1/\sqrt 2)(|0\rangle + |1\rangle)$ and
$|1\rangle \rightarrow(1/\sqrt 2)(|0\rangle - |1\rangle)$
for each qubit;
and finally, again a $1\rightarrow 3$
repetition code in the (new) $z$ basis.
All together, the encoding is:
\begin{eqnarray}
|0\rangle &\rightarrow& 
\frac{1}{\sqrt8} (|000\rangle+|111\rangle)(|000\rangle+|111\rangle)
(|000\rangle+|111\rangle) \ ,
\nonumber \\
|1\rangle &\rightarrow& 
\frac{1}{\sqrt8} (|000\rangle-|111\rangle)(|000\rangle-|111\rangle)
(|000\rangle-|111\rangle) \ . 
\label{Shor-scheme}
\end{eqnarray}
We denote it as $R_3  U R_3$ where $R_n$ 
stands for $1\rightarrow n$ 
repetition code, and $U$ for rotation from the $z$ basis to the $x$ basis.

For cryptographic purposes, one is interested in error-reduction schemes,
which leave a minimal remainder error, rather 
than in error-correction schemes, which leave a higher remainder error. 
For that aim, the majority vote decoding should be replaced by
an unanimous decision; in case of a disagreement 
the bit is thrown away.
The classical ($1\rightarrow n$) repetition code $R_n$ with $n=2 t + 1$
provides successful 
unanimous decision with probability $Q=(1-p)^n + p^n$, and the remainder
error in this case is $P=p^n/Q$ which is 
$
P\approx \left(\frac{p}{1-p}\right)^n 
$ 
for small $p$.
This code can also be used to correct up to $t$ errors, 
but with a much higher (average) remainder error, which can be calculated
from the binomial expansion of $[ p + (1-p) ]^n$.
However, if exactly 
$t$ errors were identified and corrected, the probability 
that there were actually $t+1$ errors (hence, a remainder error) is $p$.

For $n=3$ the remainder error in the error-reduction
scheme is $P\approx p^3 + 3 p^4$ which 
is much improved in 
comparison to the 
(average) remainder error in case of error-correction, and even 
the $n=2$ error-reduction code 
provides a remainder error $P \approx  p^2 + 3p^3$ which is better than 
$3 p^2 - p^3$ 
for small $p$.
This was first noted by Vaidman, Goldenberg and Wiesner (VGW)~\cite{VGW}
who presented the 
quantum error-reduction scheme
$R_2 U R_2$ to improve the remainder error
while using only 4 qubits instead of 9 for the encoding.
The error-reduction process is done by projecting the state of the code
qubits on a desirable subspace; for instance, in case of the
$n=3$ quantum error-reduction code,
it is projected on the subspace spanned by the two states
of eq.~\ref{Shor-scheme}.
If the projection fails, the
qubit is not corrected but is thrown away.
Throwing the bits has 
only small influence on a quantum key distribution protocol
since the legitimate users throw away most of the bits due to other reasons.
Note that this is not appropriate for quantum computing, 
where throwing one bit
in the computing process destroys the computation.
On the other hand,
for cryptographic purposes, the number of bits used for the decoding
is less important (in comparison to the requirements of quantum computing),
since neither of the existing protocols
makes use of the coherence of more than two particles.

Error-correction can be easily combined into an error-reduction
scheme for the price of increasing the remainder error $P$.
The benefit of such a combination is that the probability of
successful projection, $Q$, is increased.
For simplicity we shall consider only ``pure''  
error-reduction scheme,
but our scheme can be generalized to combine the correction of
few bits as well.
In a scheme which combines error-reduction and ($t'$-errors) error-correction, 
one has to check into which subspace the state is projected,
and if this subspace corresponds to $t'$ errors (or less) the state 
is corrected by simple transformations (see~\cite{Shor1} {\em etc.}).

We conclude that 
the codes which are used for quantum error-correction must be modified  
to provide error reduction in order 
to fit the requirements of quantum cryptography
much better.
For example, 
we suggest to use error-reduction codes $R_n U R_n$ with
large $n$.
Such codes encode one qubit into $N=n^2$ qubits,
in order to reduce the error-rates 
exponentially with $n$ (more efficient codes could be used as well,
based, for instance, on~\cite{Steane1,Steane2,CS}).
The rest of this paper is devoted to the analysis of these codes.
As in all discussions on quantum error-correction, coherent 
transformations of many particles are dismissed since,
in real channels, such errors are 
much smaller than errors in individual bits.
However, we consider also eavesdropping aspects, and therefore, 
this issue is more subtle and we elaborate it further later on.

It is generally
believed that it is enough to correct phase errors, bit errors
and bit-phase errors in 
order to protect against any independent error
(see the analysis in~\cite{BDSW,EM}).
The intuitive problem with such argument is the assumption that each qubit is
either strongly disturbed (due to bit flip in some basis) or not
disturbed at all, while in reality, {\em all} qubits are slightly
changed.
Following~\cite{Shor1,EM,BDSW,VGW,Steane1,Steane2} 
and other works on this 
subject we find the remainder error $P$
and the probability of success $Q$
given that 
bit errors, phase errors and phase-bit errors
occur with probability $p$. 
However, for the simple special case of the code
$R_2 U R_2$ 
we demonstrate the error reduction
explicitly by discussing a general transformation on each bit.

A qubit is described by a two-dimensional Hilbert space (say, spin of
a spin-half particle)
$ \alpha |0 \rangle + \beta | 1 \rangle $
with $|\alpha|^2 + |\beta|^2 =1$.
When it is encoded using $R_n$ we get the state 
$ | \Psi_{_{R}}\rangle = \alpha |0_{_{R}}\rangle
+ \beta |1_{_{R}}\rangle $
in a $2^n$ dimensional Hilbert space,
with $|0_{_{R}}\rangle = |0_1 0_2\cdots 0_n\rangle$
and 
$|1_{_{R}}\rangle = |1_1 1_2\cdots 1_n\rangle$.
When it is encodes using $R_nUR_n$ we get the state  
$ |\Psi_{_{RUR}}\rangle = \alpha |0_{_{RUR}}\rangle
+ \beta |1_{_{RUR}}\rangle $
of $N=n^2$ qubits in a $2^{(n^2)}$ dimensional Hilbert space,
with
\begin{eqnarray}
|0_{_{RUR}}\rangle =
\left(\frac{1}{(\sqrt2)^n}\right)
(|0_1 \cdots 0_n\rangle +
                     |1_1 \cdots 1_n\rangle) \quad \quad \quad \nonumber \\
(|0_1 \cdots 0_n\rangle +
                     |1_1 \cdots 1_n\rangle)  
\ldots 
(| 0_1 \cdots 0_n\rangle +
                     |1_1 \cdots 1_n\rangle)  
                                                     \nonumber \\ 
|1_{_{RUR}}\rangle =
\left(\frac{1}{(\sqrt2)^n}\right)
(|0_1 \cdots 0_n\rangle -
                     |1_1 \cdots 1_n\rangle) \quad \quad \quad \nonumber \\
(|0_1 \cdots 0_n\rangle -
                     |1_1 \cdots 1_n\rangle)  
\ldots 
(| 0_1 \cdots 0_n\rangle -
                     |1_1 \cdots 1_n\rangle)  
                                                    \label{my-code} 
\end{eqnarray}
where there are $n$ multiplets of $n$ bits each.
In the decoding process, the disturbed state is projected on the
desirable 2-dimensional subspace spanned by the two states
$ |0_{_{RUR}}\rangle $ and
$ |1_{_{RUR}}\rangle $.
Let us see the influence of the different types of errors on the final state.
(1) Bit errors: 
Opening the parentheses, it is easily seen that bit-errors in less 
then $n$ bits cannot bring the state back into the desirable subspace.
(2) Phase errors: It is not easy to calculate  
the number of phase errors which can bring the state back to the
relevant subspace, if we use the $z$ basis;
however, phase errors in the $z$ basis
are bit errors in the $x$ basis (see~\cite{Shor1,Shor2}, {\em etc}.).
Therefore by  
transforming this state to the $x$ basis of each qubit,
the two states 
$ |0_{_{RUR}}\rangle $ and
$ |1_{_{RUR}}\rangle $ become superposed from different words
which differ by at least $n$ bits, and thus,
$n$ phase errors are required in order to bring the original
state back to the relevant subspace. 
(3) Phase-bit errors: Showing that only $n$ such errors bring
the state back to the desirable subspace is done using the same
approach, by a transformation to the $y$ basis. 
We conclude that the probability of success and the remainder
error are indeed 
$Q \approx (1-p)^n $ and $P \approx p^n/Q$,
as calculated for the classical error reduction
scheme $R_n$.

We now provide a partial analysis
of more realistic type of errors.
Let each qubit in the code 
be transformed arbitrarily
(but independently).
In general, the transformation is not unitary since an ancila
(e.g., environment) might be involved.
However, we can still deal only with unitary transformations
and the effect of decoherence (non-unitary transformations) is 
obtained by averaging over several different 
unitary transformations with
appropriate probabilities. A similar argument is provided
in~\cite{EM}. 
Restricting ourselves to ``pure'' error-reduction schemes,
we must demand that all the individual unitary transformations be 
weak (close to unity).
In a generalization of our scheme which correct $t'$ errors,  
up to $t''$ (which is somewhat smaller than
$t'$) of the transformations are permitted not to be weak. 

We provide a complete analysis only for   
the code $R_2 U R_2$, but such analysis can 
be extended to codes $R_n U R_n$ with $n>2$. 
Let each qubit $j$ in the code  
be exposed to the most general
one-particle transformation
\begin{equation}
U_j = \left(\begin{array}{cc}
                \cos \theta_j    &  \sin \theta_j e^{ i \phi_j} \\
 - \sin \theta_j e^{ i \eta_j}  &  \cos \theta e^{ i(\phi_j + \eta_j)} 
\end{array}\right) 
\end{equation}
(up to an irrelevant overall phase),
where all angles are smaller than some small angle $\chi$,
so that $p \approx \chi^2$.
We write how the original state $|\Psi_{_{RUR}}\rangle$
in the $2^{(2^2)} = 16$ dimensional
Hilbert space is transformed (due to the 
matrix elements $\langle 0000|U_1 U_2 U_3 U_4| 0011\rangle$ {\em etc.}):
\begin{equation}
\left(\begin{array}{c} 
\alpha +\beta \\
0 \\ 
0 \\ 
\alpha -\beta \\
0 \\ 
\cdot \\ 
\alpha -\beta \\
0 \\ 
0 \\ 
\alpha +\beta \\
\end{array}\right)   \rightarrow 
\left(\begin{array}{c} 
x_{0000} \\
x_{0001} \\ 
x_{0010} \\ 
x_{0011} \\ 
x_{0101} \\ 
\cdot \\ 
x_{1100} \\ 
x_{1101} \\ 
x_{1110} \\ 
x_{1111}  
\end{array}\right) \ ,   \end{equation}
with 
$x_{0000} = (\alpha+\beta) \cos\theta_1\cos\theta_2\cos\theta_3\cos\theta_4
          + (\alpha-\beta) \cos\theta_1\cos\theta_2\sin\theta_3\sin\theta_4
                    e^{i \phi_3} e^{i \phi_4}
          + (\alpha-\beta) \sin\theta_1\sin\theta_2\cos\theta_3\cos\theta_4
                    e^{i \phi_1} e^{i \phi_2}
          + (\alpha+\beta) \sin\theta_1\sin\theta_2\sin\theta_3\sin\theta_4
                    e^{i \phi_1} e^{i \phi_2} e^{i \phi_3} e^{i \phi_4}  $
{\em etc.}
Projecting the state onto the subspace spanned by 
$ |0_{_{RUR}}\rangle = (1/2)( 
|0000\rangle + |0011\rangle + |1100\rangle + |1111\rangle)$
and
$ |1_{_{RUR}}\rangle = (1/2)( 
|0000\rangle - |0011\rangle - |1100\rangle + |1111\rangle)$,
and defining
$C=
                 (\cos \theta_1 \cdots \cos \theta_4\  \cos \phi_1
                  \cdots \cos\phi_4\  \cos \eta_1 \cdots \cos \eta_4)$,
we obtain, after a lengthy calculation, the (unnormalized)
final state
$| \Phi_{_{RUR}}\rangle = C
                \left[ {\alpha \choose \beta} + O(\chi^2) \right] $ .
The final state, when normalized, is almost identical to the initial state
$|\Psi_{_{RUR}}\rangle$,
where the terms which contribute to the correction are 
$\sin \theta_1 \sin \theta_2$; $\sin \eta_3 \sin \eta_4$, 
{\em etc.}, all of order $O(\chi^2) $ or smaller.
Thus, the remainder error probability is indeed 
$O(\chi^4) \approx O(p^2) $, 
with probability of success $C^2$.

This code can be used for $t'$-bit error-correction scheme
if we do not reject the encoded bit when the projection fails.
Instead, we check into which subspace the state is projected.
In this case the assumption that all angles are small can be dismissed, 
so that $t''$ (which is smaller than $t'$) angles can be large.
Recall however that  in these cases the remainder error-rate is not 
$O(\chi^{2n})$ anymore.

The main problem of a scheme which performs only error reduction is that 
the probability of successful projection diminishes when
$n$ is increased as $(1-p)^n$.
We could combine it with some (small-$t'$) error-correction
as previously explained, but there is also a different solution,
which should be preferable in case the noise changes in time
as
$\theta \approx wt$ {\em etc.}
In this case the probability of success can be much improved
using the Zeno effect
(see discussion in~\cite{VGW,ChuaYam}) by performing $M$ projections in between,
at equal time steps,
reducing $p$ to $p/(M^2)$, and $Q$ to $ (1-\frac{P}{M^2})^{nM}\approx
1-np/M$. 
The remainder error is also much improved by this process.
Performing $M$ projections is rather simple 
when enhancing a quantum memory is considered (meaning
that it does not add any further complication).
When transmission to long distances is considered, Alice and
Bob need to have ``projection stations'' between them.
This greatly improves $Q$ and does not affect the security of the
transmission. Indeed, since 
each such station is only required to perform the desired projection,
it can even be controlled by the eavesdropper;
if Eve tries to do anything other than the required projections ---
she increases the error-rate and will be detected.

The only assumption required for the success of any error-correction
or error-reduction scheme is that each code bit is disturbed independently 
of the others.
If real noise causes many-particle transformations
the scheme will fail, but for bits stored or transmitted separately,
such effects are
expected to be negligible.
Thus, the legitimate users of quantum cryptography can use error-reduction
schemes to decrease much of the noise, 
and, as a result, expect much less errors
when comparing a portion of the data.
It is important to note that the added assumption does not restrict
the adversary, Eve. She is still allowed to do whatever she likes,
including creating many-particle coherence.
The eavesdropping analysis needs only to take 
the final error-rate into account. 
We could even let Eve do all the transformations
from the initial bit, through the encoding till she obtains the final bit.
If she deviates from the protocols and the error-rate is larger than expected
Alice and Bob quit the transmission.
If she deviates from the protocol but provides the final state with
the allowed error-rate, Alice and Bob do not care which operations she has
done, since the allowed small error-rate (which is verified), promises
them that her information is limited as desired.
This provides a new type of a privacy amplification scheme,
simpler than the one recently suggested~\cite{England} which is based
on purification of singlets~\cite{purification}.
Such schemes provide a proof of the ultimate security of 
quantum cryptography 
under the assumption that the legitimate users have perfect devices.
Moreover, in case Eve gets the code bits without 
knowing which code bits  
encode a particular qubit, her information is 
reduced even more. 
This argument is similar to 
the {\em randomization argument}
used in~\cite{BM,England}.  It may provide the proof of perfect security
even when Alice and Bob have real devices,
since Eve cannot know which coherence would be useful to her in advance,
hence, her information is reduced whether the legitimate
users can observe this reduction or not!
However, analyzing this aspect of quantum privacy amplification
is rather complicated and it is beyond
the scopes of this work. 

In conclusion, we have shown that quantum cryptography can be much improved
by using quantum error-reduction schemes. Our result might be crucial for
implementing quantum cryptography over large scale distances and times.
It also provides an alternative quantum privacy amplification scheme.
We suggested a specific encoding which yields 
exponentially small remainder error, 
and we suggested to implement it in 
a ``many-stations'' system, so that the probability of success 
will not become too small.
The errors due to the frequent projections in a ``many-stations'' system
were not considered here. As in the case of a 
fault-tolerant calculations~\cite{Shor2}, it may well be that there is some
optimal number of stations $M$ such that a larger number of stations
causes an increase of the remainder error.
Note also that some errors are due to creation and measurement of the state
in the labs of Alice and Bob, and for the time being
these limits our ability to reduce $P$.
However, the main limitations on quantum cryptography are maintaining
coherence over long distances and times 
and these limitations are solved efficiently
using the scheme we suggest.

>From all works which recently appeared, the work of VGW~\cite{VGW} 
is more related to ours than the others.
It considers 
the use of the quantum Zeno effect and the $R_2 U R_2$ error-reduction
scheme to maintain quantum states in a
quantum memory for a longer time.  
However, this work does not deal with the benefits of using a large number of 
code bits, and with improving transmission to large distances.
Other less related works are these of~\cite{EM,BDSW} which discusses
quantum communication, and these of Steane~\cite{Steane1,Steane2}
which discusses large $n$.

The author is grateful to Gilles Brassard and 
Asher Peres for motivating this work,
and to Eli Biham, Netta Cohen, Lior Goldenberg 
and Lev Vaidman for very helpful discussions.

%\end{multicols}
\end{document}